\RequirePackage{fix-cm}
\documentclass[smallextended]{svjour3}       
\smartqed  

\usepackage{natbib}
\usepackage{graphicx}
\usepackage{amsmath,amssymb,amsfonts}
\usepackage{textcomp}
\usepackage[ruled,vlined,linesnumbered,noend]{algorithm2e}
\usepackage{graphicx}
\usepackage{caption}
\usepackage{subcaption}
\usepackage{paralist}
\usepackage{tabularx}
\usepackage{colortbl}
\usepackage{balance}
\usepackage{multirow}
\usepackage{multicol}
\usepackage{pbox}
\usepackage{mathtools}
\usepackage{algorithmic}
\usepackage{paralist}
\usepackage{url}
\usepackage{setspace}
\usepackage{verbatim}
\usepackage{float}
\usepackage[normalem]{ulem}
\usepackage{xspace}
\usepackage{pifont}
\usepackage{tikz}
\usepackage{xcolor, soul}
\usepackage[most]{tcolorbox}
\usepackage[colorlinks=true, linkcolor=black, citecolor=black]{hyperref}
\usepackage[shortlabels]{enumitem}
\usepackage{array}
\usepackage{booktabs}
\usepackage{adjustbox}
\usepackage{longtable}
\usepackage{adjustbox}
\usepackage{appendix}

\usepackage{inconsolata}
\usepackage[T1]{fontenc}
\usepackage{makecell}
\usepackage[misc]{ifsym}

\include{000_macro}

\begin{document}

\title{An empirical study of data constraint implementations in Java}

\author{Juan Manuel Florez         \and
        Laura Moreno				  \and
        Zenong Zhang				  \and
        Shiyi Wei				  \and
        Andrian Marcus
}

\institute{\Letter\xspace J. M. Florez, Z. Zhang, S. Wei, and A. Marcus \at
              Department of Computer Science \\
              The University of Texas at Dallas\\
              Richardson, TX, USA \\
              \email{\{jflorez, zenong, swei, amarcus\}@utdallas.edu}           
           \and
           L. Moreno \at
              CQSE America\\
              Santa Clara, CA, USA \\
              \email{moreno@cqse-america.com}
}

\date{ }

\maketitle

\begin{abstract}
	Software systems are designed according to guidelines and constraints defined by business rules.
	Some of these constraints define the allowable or required values for data handled by the systems. 
	These \textit{data constraints} usually originate from the problem domain (\eg regulations), and developers must write code that enforces them.
	Understanding how data constraints are implemented is essential for testing, debugging, and software change. 
	Unfortunately, there are no widely-accepted guidelines or best practices on how to implement data constraints.
	
	This paper presents an empirical study that investigates how data constraints are implemented in Java.
 	We study the implementation of \tcs data constraints extracted from the documentation of eight real-world Java software systems. 
	First, we perform a qualitative analysis of the textual description of data constraints and identify four data constraint types.
	Second, we manually identify the implementations of these data constraints and reveal that they can be grouped into \cips \textit{implementation patterns}.
	The analysis of these implementation patterns indicates that developers prefer a handful of patterns when implementing data constraints and deviations from these patterns are associated with unusual implementation decisions or code smells.
	Third, we develop a tool-assisted protocol that allows us to identify \tps additional trace links for the data constraints implemented using the 13 most common patterns. 
	We find that almost half of these data constraints have multiple enforcing statements, which are code clones of different types.

\end{abstract}

\section{Introduction}
\label{sec:introduction}

Most software systems are designed to automate processes that are described by business rules.
Business rules are therefore fundamental to the development process, as they encapsulate the knowledge that is necessary to formulate the requirements of software systems.
Eliciting and explicitly referencing business rules helps ensure that the finished software fulfills its goals \citep{Witt2012}.
Indeed, business rules have even been called ``first-class citizen[s] of the requirements world'' \citep{BusinessRulesGroup2003}.
Business rules may originate from multiple sources, and in most cases are formulated in response to external factors (e.g., policies, regulations, and industry standards) \citep{Witt2012}.
Not only is it important to correctly implement these business rules to comply with applicable regulations \citep{Rempel2014}, but the traces between the business rules and their implementations should also be made explicit to facilitate maintenance in the inevitable case where the business rules change \citep{Cemus2015, Wiegers2013, Cerny2011}.

Unfortunately, business rules are rarely thoroughly documented and traced \citep{Witt2012,Wiegers2013}.
Even when that is the case, external documentation or traces often become out of sync with other artifacts, which makes source code the only artifact that can be reliably assumed to contain this knowledge. 
This is a known open problem in software engineering and traceability \citep{Rahimi2016, Domges1998, Cleland-Huang2014a}.
A significant body of work has sought to reverse-engineer business rules from existing systems \citep{Hatano2016, Cosentino2013, Huang1996, Sneed1996, Wang2004, Cosentino2012, Sneed2001, Chaparro2012}.
Such approaches depend on developer involvement (\eg finding the relevant variables) and assumptions about how the rules are implemented.
For example, a common assumption is that rules are always implemented as conditional branches in the source code \citep{Hatano2016}.
However, these assumptions are not based on empirical evidence.
We argue that studying how developers implement business rules, identifying patterns and good practices, is important for advancing this field of research and improving software development practices.

Software engineering textbooks and research papers describe many software design and programming best practices (or anti-patterns), which are usually geared towards high-level issues (\eg system decomposition or naming conventions) or towards control and data flow organization for specific types of operations (\eg design patterns) \citep{Gamma1994,Larman2005,Fowler2018}.
In addition, companies and open source communities have their own coding standards, informed by their experiences. 
However, there are no such prescribed solutions or best practices when it comes to implementing business rules.
Existing literature offers guidance on how to formulate these rules, but not on how to implement them \citep{Wiegers2013, Witt2012}.

In this paper, we focus on analyzing the implementations of one particular type of business rules, that is, \textit{data constraints}.
A data constraint is a restriction on the possible values that an attribute may take  \citep{Wiegers2013,Witt2012}.
While all data used in a software system are subject to constraints, we focus on the constraints stemming from the business rules of the problem domain that a software system addresses. 
For example, ``\textit{[the maximum frequency] is greater than the Nyquist frequency of the wave}'' \citep{swarm:rg2} is a constraint on seismic waveform data, while ``\textit{the patient is three calendar years of age or older}'' \citep{itrust:uc51} is a constraint on healthcare data.
For simplicity, in the remainder of the paper, when we refer to \textit{constraints}, we imply \textit{data constraints}.

The study of data constraints is important because they are described in many business rules taxonomies found in the literature \citep{Wan-Kadir2004, Wiegers2013, Hay2000}.
Moreover, data constraints are common in the specifications of safety-critical systems \citep{Mader2013}.
These constraints are subject to change as business rules and regulations change.
Hence, it is essential that developers can easily (or even automatically) change, test, and verify the code implementing the constraints.

In theory, there are countless ways in which one can implement a given data constraint in a given programming language.  
However, we posit that developers, guided by their experience, are likely to converge towards an identifiable set of patterns when implementing data constraints. 
The existence of such patterns would allow the definition of best practices (in the vein of design patterns), and would support reverse engineering, traceability link recovery, testing, debugging, and code reviews, among other applications.

This paper presents an empirical study of data constraint implementations in Java. 
We extracted \tcs constraints from eight real-world open-source systems and used open coding \citep{Miles2014} to categorize them into four constraint types (\secref{sec:rq1}). 
Then, we manually traced each constraint to its implementation and categorized each into \cips \textit{data constraint implementation patterns} (\secref{sec:rq2}).
We found that \freqcips patterns are used frequently and account for the implementation of most (\ie \freqpc) constraints in our data set, while \rarecips patterns are rare (\ie used only once in the data we studied).
The data also indicate that certain patterns are used more frequently than others for different types of constraints.
In addition, we found evidence that the uses of rare patterns are likely associated with code smells and unusual implementation decisions.

Furthermore, we developed a tool-assisted protocol to identify additional statements that enforce the manually traced constraints.
This protocol is applicable to constraints that are implemented with \dets of the most frequently used implementation patterns, which cover \tdcs (\tdcscover) of the constraints in our data set.
Using this tool-assisted protocol, we recovered \tps additional statements that enforce \edscs of the \tdcs constraints.
The analysis of the new links shows that \pcmi of the \edscs constraints are enforced in at least two different parts of the code.
In most of these cases (\cconsp), the different statements use the same pattern, which indicates that developers use type 1 and type 2 code clones for these implementations (\secref{sec:rq3}).

The main contributions of our study are: 

\paragraph{(1)} A catalog and analysis of \cips data constraint implementation patterns in Java.
We expect that the catalog will support the definition of best practices and guidelines for the implementation of data constraints.
The catalog is a first step in understanding how developers implement business rules.
Our study and catalog can be a model for the study of other subsets of business rules.
We expect these implementation patterns to be essential in designing tools for supporting testing, debugging, code reviews, automatic traceability, or reverse engineering of data constraints.

\paragraph{(2)} A set of 443 curated, line-of-code-level, traceability links from \tcs data constraints definitions to their implementations, in eight real-world Java systems, which will enable future research in the area.
These links were generated partly manually, and partly using a novel tool-assisted protocol.

The implementation patterns catalog and the data used to construct is publicly available \citep{RepPack2021} for future development.

The remainder of the paper is organized as follows.
\secref{sec:motivation} introduces a motivating example, which shows and discusses the Java implementation of a particular data constraint.
\secref{sec:questions} presents the three specific research questions we address in this empirical study.
Sections \ref{sec:rq1}, \ref{sec:rq2}, and \ref{sec:rq3} describe the data, protocols and analyses we performed to answer each research question, respectively.
They also present the results and provide answers to each research question.
Section \ref{sec:threats} discusses the threats to validity and limitations of the study, while Section \ref{sec:related_work} presents the related work.
Finally, conclusions and future work are in \secref{sec:conclusions}.
The paper includes the catalog of the \tcs constraint implementation patterns as an appendix.
A subset of the most frequent ones, which fits on one page, is also included in the paper as a table, to ease reading and understanding.

\section{Motivating Example}
\label{sec:motivation}

We present the implementation of one data constraint extracted from a use case of iTrust, a healthcare system widely used in traceability research \citep{Zogaan2017}.
This use case evaluates whether a patient is at risk of suffering from type 2 diabetes according to multiple risk factors, one of them being: \textit{``Age over 45''} \citep{itrust:chronic}.
The data constraint expressed in this excerpt is $\mathit{age} > 45$.

\listref{lst:motivation} contains the code relevant to the implementation of this constraint.
In the \src{Type2DiabetesRisks} class, the \src{getDiseaseRiskFactors()} method defines and adds four risk factors in lines 4 to 7, among which we find the relevant line based on the keyword \emph{age} and the constant 45 in line 4.
The constructor of the \src{AgeFactor} class assigns the constant \src{45} to its field called \src{age}.
Examining the usages of the \src{getDiseaseRiskFactors()} method, we see that after being initialized, the \src{hasRiskFactor()} method is called on each risk factor (line 18).
This method delegates the constraint checking to the \src{hasFactor()} method. 
Finally, line 31 checks the constraint, which appears in the \src{hasFactor()} method of the \src{AgeFactor} class.

While lines 4, 18 , and 31 in \listref{lst:motivation} are all part of the implementation of the constraint, we consider that the statement that actually enforces the constraint is the last one.
We call such a statement the  \textit{constraint enforcing statement}.
For simplicity, in the remainder of the paper, when we refer to \textit{enforcing statement}, we imply \textit{constraint enforcing statement}. We provide relevant definitions in \secref{subsec:tracing}.

\begin{lstlisting}[caption={Code implementing the constraint in the motivating example},label=lst:motivation,float]
// Class Type2DiabetesRisks
protected List<PatientRiskFactor> getDiseaseRiskFactors() {
  List<PatientRiskFactor> factors = new ArrayList<>();
  factors.add(new AgeFactor(patient, 45)); // <<
  factors.add(new WeightFactor(currentHealthRecord, 25));
  factors.add(new HypertensionFactor(currentHealthRecord));
  factors.add(new CholesterolFactor(currentHealthRecord));
  return factors;
}

...

// Class RiskChecker
public boolean isAtRisk() {
  int numRisks = 0;
  List<PatientRiskFactor> factors = getDiseaseRiskFactors();
  for (PatientRiskFactor factor : factors) {
    if (factor.hasRiskFactor()) // <<
      numRisks++;
    if (numRisks >= RISK_THRESHOLD)
      return true;
  }
    
  return false;
}

...

// Class AgeFactor
public boolean hasFactor() {
  return patient.getAge() > age; // <<
}
\end{lstlisting}

This example shows that it is possible to identify a single enforcing statement for a data constraint which consists of a single expression in the code (\src{patient.getAge() > age}).
However, the data relevant to the constraint are defined in code locations different from where the constraint is being enforced.
Specifically, \src{age} is a field of class \src{Patient}, and the constant \src{45} is a parameter to the constructor call of the \src{AgeFactor} class.
This means that the enforcing statement alone is not sufficient to describe the implementation of a constraint. 
In this case, the implementation consists of (at least) the statement that enforces the constraint (\src{patient.getAge() > age}), and the definitions of \src{Person.age} and of the constant \src{45}.

We can further note that a given enforcing statement may correspond to multiple constraints, \ie any other uses of the \src{AgeFactor} class would correspond to different constraints but use the same code for enforcing them.
For example, \src{AgeFactor} initialized with the value \src{30} would check a different constraint (\ie $\mathit{age} > 30$) but would use the same code to do so.
This is a situation when \emph{multiple constraints use the same enforcing statement}.

Finally, the constraint may need to be enforced in other features of the system.
For example, the same risk factor is also used in determining whether a patient is at risk of suffering heart disease, and the \src{HeartDiseaseRisks} class contains a check for \textit{``Age over 45''}.
The implementation in this case is identical, \ie \src{AgeFactor} is initialized with the constant 45.
This is a situation when \emph{a constraint has multiple enforcing statements or uses}.

This example illustrates that, even though a constraint implementation can be traced to a single enforcing statement and corresponding data definitions, understanding how data constraints are implemented is further complicated by \emph{the need to disambiguate different constraints that use the same code}, and \emph{locating different enforcing statements of the same constraint}.
We seek to build an understanding of data constraint implementations by identifying patterns both in their textual description and their implementation.

\section{Research Questions}
\label{sec:questions}

Based on our collective experience, we posit that many unrelated constraints are implemented in similar ways.
This also implies that there should be a relatively small number of forms that constraint implementations normally take.
However, we know little about the space of data constraint implementations, which is why we conducted this empirical study.

The main goal of our study is understanding how data constraints are implemented, and we formulate three specific research questions (RQ), addressing three distinct aspects of data constraints and their implementations:

\textbf{RQ1:} \textit{What types of data constraints can be found in textual artifacts of software projects?}
For answering RQ1, we perform a qualitative analysis of the textual description of data constraints and identify the kinds of restrictions they specify (\secref{sec:rq1}).

\textbf{RQ2:} \textit{What patterns do developers follow when implementing data constraints in Java?}
For answering RQ2, we manually identify the implementations of the data constraints.
Then, we perform a qualitative analysis for identifying commonalities and differences between them (\secref{sec:rq2}).

\textbf{RQ3:} \textit{What are the differences between multiple enforcing statements of the same constraint?}
For answering RQ3, we implement a set of tools that allow us to semi-automatically identify enforcing statements additional to those identified manually before. 
Then, we analyze the multiple enforcing statements of the same constraint, when they exist, to understand their rationales (\secref{sec:rq3}).

\section{Types of Data Constraints (RQ1)}
\label{sec:rq1}

In this section, we present the data and analyses we used to answer RQ1: \textit{What types of data constraints can be found in textual artifacts of software projects?}
We then describe the results and provide the answer to the research question.

\subsection{Software Systems} 
\label{subsec:data-selection}

The targets of our empirical study are eight open-source real-world Java systems (\tabref{tab:systems}).
We selected these systems because they have been used in previous research on traceability \citep{Zogaan2017} and cover a broad range of application domains (see column 3 in \tabref{tab:systems}).
Due to the difficulty of procuring requirements documents for open source software \citep{Alspaugh2013}, we selected the textual artifacts that were available for each system, a practice common in traceability research \citep{Eaddy2008a,Ali2011,Ali2013,Ali2012a}.
These artifacts contain descriptions of the systems features and business rules.

\renewcommand{\arraystretch}{1.7}
\begin{table}[t]
	\centering
	\caption{Software systems used in the empirical study.}
	\small
	\setlength\tabcolsep{3pt}
	\setlength\extrarowheight{-2pt}
	\begin{tabular}{>{\raggedright\arraybackslash}p{2.5cm}>{\raggedright\arraybackslash}p{1.3cm}>{\raggedright\arraybackslash}p{3cm}>{\raggedleft\arraybackslash}p{1.2cm}>{\centering}p{2.5cm}}
	\toprule
		
	\multicolumn{1}{c}{\multirow{2}{*}{\textbf{System}}} & \textbf{Short} & \multicolumn{1}{c}{\multirow{2}{*}{\textbf{Domain}}} & \multicolumn{1}{c}{\textbf{Size}} & \textbf{Textual} \tabularnewline
	 & \textbf{name} &\textbf{} & \multicolumn{1}{c}{\textbf{(KLoC)}} & \textbf{artifacts} \tabularnewline
	
	\midrule

Apache Ant&Ant&Build manager&282&User Manual\tabularnewline
ArgoUML&Argo&UML Modeler&154&User Manual\tabularnewline
Guava&Guava&Programming utilities&346&User Manual\tabularnewline
Apache HTTPComponents&HTTPC&HTTP Client/Server library&28&Tutorial, Specifications\tabularnewline
jEdit&jEdit&Text editor&196&User Manual\tabularnewline
Joda-Time&Joda&Date/time library&146&User Manual\tabularnewline
Rhino&Rhino&JavaScript interpreter&77&Specifications\tabularnewline
Swarm&Swarm&Seismic wave visualizer&101&User Manual\tabularnewline

		\bottomrule

	\end{tabular}%
	\label{tab:systems}%
	\vspace{-0.2cm}
\end{table}

\subsection{Constraint Extraction and Categorization}

One author extracted the data constraints from the textual artifacts of the target systems (see column 5 in \tabref{tab:systems}).
A total of 198 constraints were extracted, out of which 11 were discarded after a discussion during the categorization (see below), leaving \tcs constraints to be used for the study.
We required the constraints to be explicit, \ie the attributes and values needed to be stated in the text.
An example is the sentence ``\textit{By default the pool allows only 20 concurrent connections}'' \citep{httpc:tutorial}, where the attribute \emph{pool} and the value 20 are explicitly stated.
An example of a constraint where not all values are explicit is: ``\textit{The default [compiler] is javac1.x with x depending on the JDK version you use while you are running Ant}'' \citep{ant:javac}.
While this sentence describes a constraint, the value of \emph{x} is not stated in the text, as it varies depending on the execution environment.
The number of constraints we identified varies across systems (see column 6 in \tabref{tab:typ-vs-sys}), because some of the artifacts define fewer data constraints than others.

We categorized all extracted constraints using open coding \citep{Miles2014}.
Each constraint was categorized according to how many operands it included and what kind of operations it applied to them.
The coding was performed by two authors.
Each coder independently coded each constraint, with the other coder verifying each other's work afterwards.

The coders had disagreements on 11 of the labeled sentences, and after a discussion, it was decided that these were not valid data constraints.
For example, the Guava manual contains the sentence \textit{``If your cache should not grow beyond a certain size''} \citep{guava}, which may appear to be a constraint but instead  describes a user consideration that is outside of the scope of the system.

\subsection{Results}
\label{sec:rq1-results}

\renewcommand{\arraystretch}{0.92}
\begin{table*}[htbp]
	\centering
	\caption{Data constraint types.\vspace{-0.3cm}}
	
	\small
	
	\begin{tabular}{p{\textwidth}}
		
		\toprule
		
		\textbf{Name:} \tVC. \\
		\textbf{Definition:} The value of an attribute $X$ is constrained by the value of another attribute $Y$ (or constant $C$). Equality or relational operators are used to determine if $X$  is greater than, less than, equal to or not equal to $Y$ (or $C$). While equality operators apply to values of all types, relational operators do not apply to all types of values (\eg binary or categorical).\\
		\textbf{Example constraint text:} ``\textit{While SWARM will allow the maximum frequency to be set to any positive value greater than the minimum frequency, this value will adjust automatically if it is greater than the Nyquist frequency of the wave being manipulated.}'' \citep{swarm:rg2}\\ 
		\textbf{Example constraint(s):} \\
		\quad $\mathit{max\,frequency} > 0$ \\ 
		\quad $\mathit{max\,frequency} > \mathit{min\,frequency}$ \\
		\quad $\mathit{max\,frequency} > \mathit{wave\,Nyquist\,frequency}$\\
		\midrule	
		
		\textbf{Name:} \tDVC. \\
		\textbf{Definition:} An attribute has only two possible, mutually-exclusive values (\eg true/false, on/off, null/not-null). Equivalent to \tVC if the operator is equality and there are only 2 possible values.\\
		\textbf{Example constraint text:} ``\textit{If configuration file is not available or readable it will default to  'UTC'.}'' \citep{swarm:rg2}\\ 
		\textbf{Example constraint(s):} \\
		\quad $\mathit{file}\, \mathit{available}==\mathtt{false}$\\
		\quad $\mathit{file}\, \mathit{readable}==\mathtt{false}$\\
		
		\midrule	
		
		\textbf{Name:} \tCTV. \\
		\textbf{Definition:} The value of a categorical attribute is constrained to a finite set of values. \\
		\textbf{Example constraint text:} ``\textit{A target has the following attributes: [...] onMissingExtensionPoint: What to do if this target tries to extend a missing extension-point. (fail, warn, ignore)}'' \citep{ant:targets}\\ 
		\textbf{Example constraint(s):} \\
		\quad $\mathit{onMissingExtensionPoint} \in \lbrace \mathit{fail}, \mathit{warn}, \textit{ignore}\rbrace $\\
		
		\midrule	
		
		\textbf{Name:} \tCV. \\
		\textbf{Definition:} The constraint directly dictates what value the property should have. \\
		\textbf{Example constraint text:} ``\textit{The GregorianJulian calendar is a combination of two separate calendar systems, the Gregorian and the Julian. The switch from one to the other occurs at a configurable date. The default date is 1582-10-15, as defined by Pope Gregory XIII}.'' \citep{jodatime:gjcal}\\ 
		\textbf{Example constraint(s):} \\
		\quad $\mathit{switch\,date}\;\mathrm{is}\; \mathit{1582}$-$\mathit{10}$-$\mathit{15}$\\
		
		\bottomrule
	
	\end{tabular}
	
	\label{tab:c-types}%
\end{table*}%

We identified four data constraint types: \tVC, \tDVC, \tCTV, and \tCV.
\tabref{tab:c-types} defines each type and provides examples.

We encountered constraints of different types in each system, yet not all constraint types appeared in the artifacts of all systems.
The distribution across types and systems of the \tcs constraints is presented in \tabref{tab:typ-vs-sys}.

\begin{table*}[htbp]
	\centering
	\caption{Distribution of constraint types by system.}
	\small
	\setlength\tabcolsep{3.5pt}
	\setlength\extrarowheight{-2pt}
	\begin{tabular}{lccccc}
	\toprule
		
	\multirow{2}{*}{\textbf{System}} & \thl{Categorical} & \thl{Concrete} & \thl{Dual Value} & \thl{Value} & \multirow{2}{*}{\textbf{Total}} \tabularnewline
	
	 & \thl{Value} & \thl{Value} & \thl{Comparison} & \thl{Comparison} & \tabularnewline
	
	\midrule

Ant&6&7&18&1&32\tabularnewline
Argo&3&14&11&$\cdot$&28\tabularnewline
Guava&$\cdot$&$\cdot$&8&3&11\tabularnewline
HTTPC&1&3&12&5&21\tabularnewline
jEdit&1&3&18&11&33\tabularnewline
Joda&2&5&$\cdot$&6&13\tabularnewline
Rhino&3&1&6&12&22\tabularnewline
Swarm&2&5&12&8&27\tabularnewline
\midrule
\textbf{Total}&18&38&85&46&187\tabularnewline

	\bottomrule
	 \tabularnewline
		
	\end{tabular}%
	\label{tab:typ-vs-sys}%
	\vspace{-0.2cm}
\end{table*}	

The most common constraint type in our data set is \tVC, in which two values are compared using an operator.
\tDVC is a subtype of \tVC, where the operator is equality, and the property can only take one of two mutually-exclusive values.
This subtype is important, because often only one of the two mutually-exclusive values is explicit in the constraint description, but it is easy to infer the missing one. 
The same inference is not possible for constraints of the more generic type, \tVC.
\tCV directly states the value that an attribute should have.
Finally, \tCTV does not specify or compare a specific value. Instead, it restricts the value of an attribute to a finite set of items.
Note that this last type only ensures that the value is an item of the given set.
A constraint requiring that a value is equal to \emph{one} of the items in the set would instead be of the \tVC type.

\rqa{\bf{RQ1 answer}}{
	We identified four data constraint types: \tVC, \tDVC, \tCTV, and \tCV. 
	They differ from one another by the number of operands they include and the type of operations applied to them.
}

\section{Constraint Implementation Patterns (RQ2)}
\label{sec:rq2}

In this section, we describe the data, protocols, and analyses we used for answering RQ2: \textit{What patterns do developers follow when implementing data constraints in Java?}
We then describe the results and provide the answer to the research question.

\subsection{Manual Tracing Protocol} 
\label{subsec:tracing}

Answering this research question requires identifying the implementation of the \tcs constraint that we extracted in the previous section.
We borrow terminology from software traceability  research  and call this activity \emph{tracing}.
This type of tracing is common in requirements-to-code traceability link recovery and feature location work, among others \citep{DeLucia2012, Razzaq2018}.
Consequently, a \emph{trace} is a link between the description of a constraint (\ie its source) and the code that implements it (\ie its target).
Tracing was performed by six Computer Science graduate students: five M.S. students, with at least two years of industry experience each, and one Ph.D. student. 
We refer to the six students as \textit{tracers} from this point forward.

Each tracer received one hour of training from one of the authors, and was compensated with \$15 per hour for the time spent in training and tracing.
The tracers worked at home, using an online spreadsheet to record their traces.
Each trace consisted of its source constraint (\ie the constraint description) and its target code statements (\ie the enforcing statements and data definitions), whose identification protocol is described further in this section.

The tracers did not communicate with one another.
Each constraint was traced by two tracers independently, and tracing proceeded one system at a time.

For each system, the tracers received the following data:
\begin{enumerate}
\item The source code of the system.

\item A document with details about the system design and architecture, such as a list of the most important classes and their responsibilities. This was assembled by one author according to the documentation of the system and a code inspection.

\item The list of  data constraints to be traced.
\end{enumerate}

In addition, for each constraint, we provided the tracers with:
\begin{enumerate}
\item The section of the textual artifact where the constraint is described, \eg a section of the user manual or specification.

\item The text that describes the constraint, \eg ``\textit{Any Content-Length greater than or equal to zero is a valid value}'' \citep{http:spec}.

\item A simplified version of the constraint, \eg ``\textit{Content-Length $\geq$ 0}''. 
This was created by one of the authors, who rephrased the textual description of the constraint as a mathematical expression.
This information was provided to ease understanding of the constraint and avoid confusion or ambiguity.

\item A scenario to be used for tracing. The scenario corresponds to a feature of the system that relies on the constraint. It was extracted by one of the authors based on the constraint's context, \eg \textit{``Validating an HTTP request''} for the example above.
This information ensures that the relevant implementation is found, as a single constraint may have multiple enforcing statements, corresponding to different features or scenarios (see \secref{sec:motivation}).
\end{enumerate}

The tracers were allowed to use any tool or information source to perform the tracing, although the use of an IDE was recommended.

\subsubsection{Structure and granularity of constraint implementations}
When tracing domain level concepts to their implementation in the code, one important aspect to establish is the granularity of the links (\ie the source and the target).
As mentioned above, our source corresponds to the textual description of a single data constraint, typically expressed in a sentence.
We discuss here the structure and granularity of the target (\ie source code elements) of the traces.

Existing work on traceability link recovery and feature location usually links sources to functions, methods, classes, files, \etc \citep{DeLucia2012}.
In other words, they use the granularity provided by the file system or the decomposition mechanism of the program language.
For our study, such a granularity is not suitable.
Recall that our goal is to study \textit{how} data constraints are implemented in Java.
For example, determining that the constraint \textit{``Age over 45''} is implemented in class \src{AgeFactor} will tell us \textit{where} it is implemented but not \textit{how}.
We need finer-grained traces (\ie to line-of-code level) to analyze and understand \textit{how} the constraints are implemented.
Conversely, as evidenced by the motivating example (\secref{sec:motivation}), tracing a constraint to a single enforcing statement or expression can be ambiguous (in the case that the same code is used to enforce multiple constraints).
We aimed to identify the minimum number of statements that unambiguously correspond to a given constraint in its context (\ie the associated feature).

For this reason, we instructed the tracers to locate both the \emph{constraint enforcing statement} and the \emph{data definition statements} for each constraint, as the tracing targets.
\emph{Constraint enforcing statements} check the constraint (\eg \src{patient.getAge() > age}) or ensure that it is enforced (\eg by directly defining the value), while \emph{data definition statements} define the data used therein (\eg \src{Person.age}, \src{45}).
We provided additional instructions for helping the tracers identify these statements.

\subsubsection{Identifying the constraint enforcing statements}

As we saw in our motivating example from \secref{sec:motivation}, several statements may be used for implementing a constraint.
Among them, we consider the enforcing statement the one that is at the lowest granularity level, that is, it cannot be decomposed any further (\eg tracing into a method invoked from the statement). 
Specifically, the tracers applied the following procedure for identifying the enforcing statements from those that are involved in implementing the constraint.
Let $s$ be a candidate enforcing statement for a given constraint:

\begin{enumerate}
	\item If $s$ contains no method invocation, then $s$ is a constraint enforcing statement. Otherwise, investigate the method $M$ invoked from $s$.
	\item If a candidate enforcing statement $s'$ exists in $M$, then repeat step (1) with $s'$. Otherwise, $s$ is a constraint enforcing statement.
\end{enumerate}

For example, when tracing the constraint \textit{``[Spectrogram maximum frequency] is greater than the Nyquist frequency of the wave''} \citep{swarm:rg2}, the tracer finds the call to \src{processSettings()}.
Inside this method, there exists the statement  \src{if(settings.spectrogramMaxFreq > wave.getNyquist())}.
The call to \src{processSettings()} is not the enforcing statement, because there is another candidate enforcing statement inside the method.
Since no statement in the method \src{getNyquist()} checks that the max frequency is greater than the Nyquist frequency, \src{if(settings.spectrogramMaxFreq > wave.getNyquist())} is the enforcing statement for the constraint in this example.

The enforcing statements should be in the code of the target system, rather than in the code of third-party libraries or the Java standard library. 
If a constraint was enforced outside the system code, tracers were instructed to trace to the statement(s) that referred to the external enforcement (\ie an invocation to the library method).
For example, the constraint ``\textit{the Iterable is empty}'' \citep{guava:prec} is implemented in the statement \src{E minSoFar = iterator.next();}.
The checking is done inside the \src{next()} method, which is implemented in the Java standard library.

\subsubsection{Identifying the data definition statements}
Using the enforcing statements, the tracers found the data definition statements by identifying the operands relevant to the constraint. 
Following the previous example, if the enforcing statement is \src{if(settings.spectrogramMaxFreq > wave.getNyquist())}, then the operands relevant to this constraint are the return value of the method \src{getNyquist()} and the field \src{spectrogramMaxFreq}.
The tracers were asked to trace the data definition statements according to the following rules:

\begin{itemize}
	\item If the data is accessed from a field directly (\src{obj.field}) or a getter method that returns the value unchanged (\src{obj.getField()}), then the \textit{field declaration} is traced.
	
	\item If the data is computed in a method, then the \textit{method declaration} is traced. For example, if the operand is the return value of the method \src{obj.calculateValue()} and
	\src{value} does not exist as a field in the class of \src{obj}, then the data definition statement is the declaration of the \src{calculateValue()} method.
	
	\item If the data comes from a library class, then the \textit{method call} is traced. 
	For example, if the operand is the local variable \src{value} which is defined as \src{int value = request.getValue()},
	where \src{request} is an instance of a library class, the data definition statement is the statement where \src{request.getValue()} is called.
	
	\item If the data corresponds to a literal defined in a method, then the \textit{assignment} is traced. 
	For example, if the operand is the variable \src{value} defined as \src{int value = 100}, this statement is traced.
	If the literal \src{100} is used directly, the data definition is the value \src{100}.
	
	\item If the constraint directly refers to a method parameter (\eg a library entry point), then the \textit{method parameter definition} is traced. 
	For example, the constraint ``\textit{the value is not null}' \citep{guava:prec} refers specifically to the parameter of the \src{checkNotNull()} method.
	This also applies when it is not possible to directly determine the caller of the method where the enforcing statement is located, \eg when the method implements a listener interface called by the \textit{Swing} library.
	
\end{itemize}

\subsubsection{Trace validation}
\label{subsec:trace-validation}
The final traces were decided jointly by two authors.
The authors applied their knowledge of the systems and the definitions presented in this section to judge whether or not the semantics of the statements marked by the tracers correspond to the constraint implementation.
This kind of approach has been applied in traceability studies when it is not possible to seek the guidance of the developers of the system \citep{Eaddy2008a, Eaddy2008}.

If the two tracers produced overlapping statements for a constraint, the trace was defined by the overlapping statements, once confirmed by one of the authors.
Disagreements (\ie the tracers identified disjoint sets of statements) were resolved by one author, with the trace being later verified and validated by another author.
Overall, only 13 (7\%) traces resulted in disagreements that were resolved through the discussion between one author and the tracers.
Some of the disagreements were caused by the misunderstandings of the code semantics (\eg the variable being checked is related, but it is not the one that the constraint refers to).
In other cases, the enforcing statement was correct but did not match the scenario outlined in the documentation.

In the end, each of the \tcs constraints was traced to one enforcing statement and the corresponding data definition statements.
As we discussed in \secref{sec:motivation}, some constraints may be involved in several features of the system, which may lead to multiple traces.
Here, we focused on providing a single trace per constraint, so we produced \tcs traces.

\subsection{Identifying Patterns in Constraint Implementations}
\label{subsec:patterns}

While the \tcs constraint implementations we traced are different from one another, they share structural properties.
We grouped them into categories according to structural properties they share.
We used open coding \citep{Miles2014} to define these categories, based exclusively on the data (\ie a descriptive approach as opposed to a prescriptive one).

\subsubsection{Coding Protocol}
Open coding results in the creation of a set of \emph{codes}, which we denominate \emph{constraint implementation patterns} (CIPs).
From here on, we refer to them as \textit{patterns} or CIPs.

In order to determine an initial set of codes (\ie patterns), we conducted an exploratory study on iTrust \citep{Zogaan2017}.
One of the authors extracted 110 data constraints from iTrust's requirements and traced them to their implementations, in a similar manner to the protocol described above.
Upon analysis of the traces, the author identified 27 patterns, which served as the initial set for the coding.
Note that the iTrust traces are not included in this study.

The CIPs have the following components:
\textit{name}, \textit{description}, \textit{statement type}, \textit{parts}, and \textit{example}.
The \textit{name} and \textit{description} of a pattern support identifying and understanding the meaning of each pattern.
The pattern's \textit{statement type} describes the type of the enforcing statement (\eg expression, method  call, or variable definition), and each pattern is defined only on a particular \textit{statement type}.
The pattern's \textit{parts} are structural programming elements from the enforcing statement. 
These \textit{parts} describe number and types of the operands and operators and differentiate one pattern from another.

\renewcommand{\arraystretch}{0.92}
\begin{table*}[htbp]
			
	\caption{7 most frequently used constraint implementation patterns form our CIP catalog. The remaining 23 are available in an appendix and the replication package \citep{RepPack2021}.}
	\label{tab:CIP}%
	
	\small
	
	\begin{tabular}{p{\textwidth}}
		
		\toprule
		
		\textbf{CIP name:} \pBP. \\
		\textbf{Description:} A variable of type Boolean is checked in a Boolean expression.\\
		\textbf{Statement type:} Boolean expression. \\
		\textbf{Parts:} $\{\mathit{variable}\}$ \\ 
		\textbf{Example:} \\ 
		\quad \textit{Instance:} \srcf{if(buffer.isModified)} \\
		\quad \textit{Parts:} $\{$\srcf{isModified}$\}$.\\
		\midrule	
		
		\textbf{CIP name:} \pBC. \\
		\textbf{Description:} Two variables are compared using one of the relational operators(\textit{equals}, \textit{does not equal}, \textit{greater than}, \textit{less than}, \textit{greater than or equal to}, \textit{less than or equal to}). Use of the \src{equals} method is considered an operator in this case. The types of the operands may be of any type for which these operations are allowed (\eg \textit{greater than} cannot be applied to Boolean values). \\
		\textbf{Statement type:} Relational expression. \\
		\textbf{Parts:} $\{\mathit{variable 1}, \mathit{op} \in \{>, \geq, <, \le, =, \ne\}, \mathit{variable 2}\}$ \\
		\textbf{Example:} \\ 
		\quad \textit{Instance:} \srcf{if(maxFreq > wave.getNyquist())} \\
		\quad \textit{Parts:} $\{$\srcf{maxFreq}, $>$, \srcf{wave.getNyquist()}$\}$.\\
		\midrule	
		
		\textbf{CIP name:} \pCA. \\
		\textbf{Description:} A literal value is passed as a parameter to a method call.\\
		\textbf{Statement type:} Method call. \\
		\textbf{Parts:} $\{\mathit{method}, \mathit{constant}\}$ \\ 
		\textbf{Example:} \\ 
		\quad \textit{Instance:} \srcf{settings.setShowVisibilities(false);} \\
		\quad \textit{Parts:} $\{$\srcf{setShowVisibilities}, \srcf{false}$\}$.\\
		\midrule	
		
		\textbf{CIP name:} \pNC. \\
		\textbf{Description:} A nullable variable is tested for nullity using the \texttt{\footnotesize ==} or \texttt{\footnotesize !=} operators. The \src{null} keyword is not considered a part of the pattern because it will always appear in instances of it.\\
		\textbf{Statement type:} Relational expression. \\
		\textbf{Parts:} $\{\mathit{variable}\}$ \\ 
		\textbf{Example:} \\ 
		\quad \textit{Instance:} \srcf{if(name == null)} \\
		\quad \textit{Parts:} $\{$\srcf{name}$\}$.\\
		\midrule	
		
		\textbf{CIP name:} \pAC. \\
		\textbf{Description:} A literal value is assigned to a variable.\\
		\textbf{Statement type:} Assignment. \\
		\textbf{Parts:} $\{\mathit{variable}, \mathit{constant}\}$ \\ 
		\textbf{Example:} \\ 
		\quad \textit{Instance:} \srcf{refreshInterval = 15;} \\
		\quad \textit{Parts:} $\{$\srcf{refreshInterval}, \srcf{15}$\}$.\\
		\midrule	
		
		\textbf{CIP name:}\pBFC. \\
		\textbf{Description:} An integer variable is used as a bit field and checked with a bitwise operator against a constant integer. \\
		\textbf{Statement type:} Relational expression. \\
		\textbf{Parts:} $\{\mathit{variable}, \mathit{constant}\}$ \\ 
		\textbf{Example:} \\ 
		\quad \textit{Instance:} \srcf{flag \& NEW\_FILE == NEW\_FILE} \\
		\quad \textit{Parts:} $\{$\srcf{flag, NEW\_FILE}$\}$.\\
		\midrule
	
		\textbf{CIP name:} \pIC. \\
		\textbf{Description:} A chain of ifs is used like a switch on a variable, checking against its possible values. Each \src{if} clause uses the \texttt{\footnotesize ==} operator or \src{equals} method.\\
		\textbf{Statement type:} If statement. \\
		\textbf{Parts:} $\{\mathit{variable}\}$ \\ 
		\textbf{Example:} \\ 
		\quad \textit{Instance:} \srcf{if(onset == EMERGENT) \{$\ldots$\} else if(onset == IMPULSIVE) \{$\ldots$\} else if $\ldots$} \\
		\quad \textit{Parts:} $\{$\srcf{onset}$\}$.\\
		
		\bottomrule	
	\end{tabular}
\end{table*}

The \textit{statement type} and \textit{parts} derived from an enforcing statement determine how to label it (\ie which patterns it follows).
For example, the enforcing statement \src{maxFrequency > nyquistFrequency} is implemented using the \pBC  pattern, because it is an expression and \src{maxFrequency}, \src{>}, and \src{nyquistFrequency} match the \textit{parts} $\{\mathit{variable 1}, \mathit{op} \in \{>, \geq, <, \le, =, \ne\}, \mathit{variable 2}\}$ respectively (see Table \ref{tab:CIP}).
Finally, the examples provide an illustration of each pattern.

Two authors coded the \tcs traces from \secref{subsec:tracing}.
Each enforcing statement was categorized according to (1) what type of statement it is, and (2) the number and types of operands and operators involved in it.
If no existing pattern matched the type of the enforcing statement or the amount of operands and operators, a new pattern was created.
As the coding progressed, patterns were renamed and/or merged, and the previously coded data were re-checked against the new CIPs.

Each trace was coded by one author.
The coded trace was verified by the other coder, discussing any disagreements with the original coder.
This is an adaptation of \textit{gold-standard coding}, in which two coders evenly split the data set, with additional \textit{reliability coders} verifying the work.
In our case, each coder acted as each other's reliability coder \citep{Syed2015}.
In 16 (9\%) cases, there were disagreements that were resolved through discussions.

\subsection{Results and Analysis} 
\label{subsec:RQ1results}

The open coding resulted in the definition of \cips CIPs. 
We organized them into a catalog of CIPs, which we present, analyze and discuss below.

\subsubsection{CIP Catalog}
Table \ref{tab:CIP} shows part of the catalog, containing the 7 most commonly used CIPs.
The complete CIP catalog, including all \cips identified CIPs, is included as an appendix and also in our replication package \citep{RepPack2021}.
Four of the \tcs (2\%) constraint implementations rely on external libraries. 
While we traced these constraints to the relevant library method call (as explained above), they were not used in defining the \cips patterns.
For this reason, we limit the following analysis to \tcsne constraints (\ie excluding the 4 enforced externally).

\renewcommand{\arraystretch}{1.3}
\begin{table*}[htbp]
	\centering
	\caption{Distribution of pattern instances by system.}
	\footnotesize
	\setlength\tabcolsep{1pt}
	\setlength\extrarowheight{-2pt}
	\begin{tabular}{p{3.5cm}rrrrrrrrr}
	\toprule
		
	\textbf{Pattern} & \textbf{Ant} & \textbf{Argo} & \textbf{Guava} & \textbf{HTTPC} & \textbf{jEdit} & \textbf{Joda} & \textbf{Rhino} & \textbf{Swarm} & \textbf{Total} \tabularnewline
	
	\midrule

\pBP$^*$&12&8&3&3&10&$\cdot$&2&11&49\tabularnewline
\pBC$^*$&2&$\cdot$&3&8&12&6&8&7&46\tabularnewline
\pCA$^*$&2&9&$\cdot$&1&$\cdot$&5&1&2&20\tabularnewline
\pNC$^*$&3&$\cdot$&2&5&1&$\cdot$&3&1&15\tabularnewline
\pAC$^*$&3&1&$\cdot$&1&1&$\cdot$&$\cdot$&3&9\tabularnewline
\midrule
\pIC$^*$&1&3&$\cdot$&$\cdot$&$\cdot$&$\cdot$&$\cdot$&1&5\tabularnewline
\pBFC$^*$&$\cdot$&$\cdot$&$\cdot$&$\cdot$&2&1&2&$\cdot$&5\tabularnewline
\pEOC$^*$&3&$\cdot$&$\cdot$&$\cdot$&$\cdot$&$\cdot$&1&$\cdot$&4\tabularnewline
\pPF&$\cdot$&1&$\cdot$&$\cdot$&2&$\cdot$&$\cdot$&$\cdot$&3\tabularnewline
\pSLC$^*$&$\cdot$&$\cdot$&$\cdot$&$\cdot$&$\cdot$&$\cdot$&2&$\cdot$&2\tabularnewline
\midrule
\pSC$^*$&$\cdot$&$\cdot$&$\cdot$&$\cdot$&$\cdot$&$\cdot$&2&$\cdot$&2\tabularnewline
\pRC$^*$&$\cdot$&$\cdot$&$\cdot$&2&$\cdot$&$\cdot$&$\cdot$&$\cdot$&2\tabularnewline
\pPM&$\cdot$&$\cdot$&2&$\cdot$&$\cdot$&$\cdot$&$\cdot$&$\cdot$&2\tabularnewline
\pNZC$^*$&$\cdot$&$\cdot$&$\cdot$&$\cdot$&2&$\cdot$&$\cdot$&$\cdot$&2\tabularnewline
\pNEC$^*$&1&1&$\cdot$&$\cdot$&$\cdot$&$\cdot$&$\cdot$&$\cdot$&2\tabularnewline
\midrule
\pSwC&$\cdot$&$\cdot$&$\cdot$&$\cdot$&1&$\cdot$&$\cdot$&$\cdot$&1\tabularnewline
\pSS&$\cdot$&$\cdot$&$\cdot$&$\cdot$&1&$\cdot$&$\cdot$&$\cdot$&1\tabularnewline
\pSE&$\cdot$&1&$\cdot$&$\cdot$&$\cdot$&$\cdot$&$\cdot$&$\cdot$&1\tabularnewline
\pS&1&$\cdot$&$\cdot$&$\cdot$&$\cdot$&$\cdot$&$\cdot$&$\cdot$&1\tabularnewline
\pOVS&$\cdot$&1&$\cdot$&$\cdot$&$\cdot$&$\cdot$&$\cdot$&$\cdot$&1\tabularnewline
\midrule
\pNBC&$\cdot$&1&$\cdot$&$\cdot$&$\cdot$&$\cdot$&$\cdot$&$\cdot$&1\tabularnewline
\pMO&$\cdot$&$\cdot$&$\cdot$&$\cdot$&$\cdot$&1&$\cdot$&$\cdot$&1\tabularnewline
\pICL&1&$\cdot$&$\cdot$&$\cdot$&$\cdot$&$\cdot$&$\cdot$&$\cdot$&1\tabularnewline
\pILF&1&$\cdot$&$\cdot$&$\cdot$&$\cdot$&$\cdot$&$\cdot$&$\cdot$&1\tabularnewline
\pIRC&1&$\cdot$&$\cdot$&$\cdot$&$\cdot$&$\cdot$&$\cdot$&$\cdot$&1\tabularnewline
\midrule
\pEV&$\cdot$&$\cdot$&$\cdot$&$\cdot$&1&$\cdot$&$\cdot$&$\cdot$&1\tabularnewline
\pDC&$\cdot$&$\cdot$&$\cdot$&1&$\cdot$&$\cdot$&$\cdot$&$\cdot$&1\tabularnewline
\pCAs&$\cdot$&1&$\cdot$&$\cdot$&$\cdot$&$\cdot$&$\cdot$&$\cdot$&1\tabularnewline
\pCSC&$\cdot$&$\cdot$&$\cdot$&$\cdot$&$\cdot$&$\cdot$&1&$\cdot$&1\tabularnewline
\pACC&1&$\cdot$&$\cdot$&$\cdot$&$\cdot$&$\cdot$&$\cdot$&$\cdot$&1\tabularnewline
\midrule
\textit{external}&$\cdot$&1&1&$\cdot$&$\cdot$&$\cdot$&$\cdot$&2&4\tabularnewline
\midrule
\textbf{Total}&32&28&11&21&33&13&22&27&187\tabularnewline

		\bottomrule
		\tabularnewline
		\multicolumn{10}{r}{$^*$Pattern that has a detector for the tool-assisted study in \secref{sec:rq3}.} \tabularnewline
		
	\end{tabular}%
	\label{tab:cip_distr}%
	\vspace{-0.2cm}
\end{table*}

\tabref{tab:cip_distr} shows the distribution of pattern instances across systems for all CIPs.
Out of the \cips patterns, \freqcips (\freqcipspc) are used to implement \freqpcimp (\freqpc) of the \tcsne constraints in our data, and we consider them \textit{frequent patterns.}
\rarecips of the patterns have only one instance in our data and we consider them \textit{rare patterns}.
The two most common patterns (\ie \pBC and \pBP) appear in nearly every system, and they alone account for 50\% of all constraint implementations in our data.
We consider these \textit{very frequent patterns}.

\subsubsection{Catalog Analysis}

\paragraph{Rare patterns.} We examined the rare patterns and those that appear in only one system and found that they usually represent implementations that stem from specific coding standards, system architecture, or developer preferences. 
Mind that while they occur rarely in our data set, we cannot claim that these implementations are unique to these systems.
In other words, studying more constraint implementations from more systems may reveal more instances of these implementation patterns. 

One example of a rare pattern is the implementation of the constraint ``\textit{[Call to ToNumber] is NaN}'' \citep{rhino:ecma}, which is simplified as ``\textit{ResultOfToNumber == NaN}''. 
An intuitive way of implementing this constraint would be \src{d == Double.NaN} (\ie an instance of the \pBC pattern). 
However, in Java, the value of \src{NaN} is not equal to itself, which leads to the implementation \src{d != d}.
We call this pattern \pSC.

A somewhat similar situation happens with the \pSLC pattern (only used in Rhino), which is the result of an optimization specific to the Rhino system, arising from a need to make the code more efficient.
According to the developers of the system: ``It is used in every native Rhino class that needs to look up a property name from a string, and does so more efficiently than a long sequence of individual string comparisons''.\footnote{\url{https://groups.google.com/g/mozilla-rhino/c/bdEX2Wa3pSQ/m/QXizSSdGEwAJ}}.
They further explain that the pattern was first devised in older versions of Java where it made a significant difference in performance, though it is not clear if that still is the case.

\begin{table*}[htbp]
	\centering
	\caption{Distribution of pattern instances by constraint type for frequent patterns.}
	\small
	\setlength\tabcolsep{3.5pt}
	\setlength\extrarowheight{-2pt}
	\begin{tabular}{>{\centering}m{3.5cm}ccccc}
	\toprule
		
	\multirow{2}{*}{\textbf{Pattern}} & \thl{Categorical} & \thl{Concrete} & \thl{Dual Value} & \thl{Value} \tabularnewline
	
	 & \thl{Value} & \thl{Value} & \thl{Comparison} & \thl{Comparison}  \tabularnewline
	
	\midrule

\pBP&$\cdot$&$\cdot$&48&1\tabularnewline
\pBC&1&$\cdot$&6&39\tabularnewline
\pCA&$\cdot$&20&$\cdot$&$\cdot$\tabularnewline
\pNC&$\cdot$&$\cdot$&14&1\tabularnewline
\pAC&$\cdot$&8&1&$\cdot$\tabularnewline
\pIC&5&$\cdot$&$\cdot$&$\cdot$\tabularnewline
\pBFC&1&$\cdot$&4&$\cdot$\tabularnewline
\pEOC&3&$\cdot$&1&$\cdot$\tabularnewline
\pPF&$\cdot$&3&$\cdot$&$\cdot$\tabularnewline
\pSLC&2&$\cdot$&$\cdot$&$\cdot$\tabularnewline
\pSC&$\cdot$&$\cdot$&$\cdot$&2\tabularnewline
\pRC&$\cdot$&2&$\cdot$&$\cdot$\tabularnewline
\pPM&$\cdot$&$\cdot$&2&$\cdot$\tabularnewline
\pNZC&$\cdot$&$\cdot$&2&$\cdot$\tabularnewline
\pNEC&$\cdot$&$\cdot$&2&$\cdot$\tabularnewline
\midrule
\textbf{Total}&12&33&81&43\tabularnewline

	\bottomrule
	 \tabularnewline
		
	\end{tabular}%
	\label{tab:cip-vs-typ}%
	\vspace{-0.2cm}
\end{table*}

\paragraph{Relationship between constraint types and implementation patterns.} The data in \tabref{tab:cip-vs-typ} indicate that there is a correlation between certain constraint types and patterns.
For example, that the two most common patterns, \pBP and \pBC implement mostly constraints of types \tDVC and \tVC, respectively.
This indicates that such constraints are implemented in rather predictable ways.
It can be argued that \pBP (\ie checking a Boolean) is an intuitive way of checking that a variable can only take two values.
Likewise, \pBC (\ie comparing two variables) is an intuitive way of implementing a comparison of two values.

Another example is the \tCTV type, which checks whether a value belongs to a finite set of options, and it is most frequently implemented with the \pIC pattern, as a series of \src{if} statements.
One can argue for using a \src{switch} statement instead. 
However, none of the implementations we examined checked constraints of the \tCTV type in this way.
Of course, semantically a \src{switch} statement is equivalent with a chain of \src{if} statements, but they differ structurally.

In the case of the \tDVC type constraints, we find a second common implementation using the \pNC pattern.
In this case, instead of checking a boolean variable, a nullable variable is checked for presence, with present/absent being the two possible values.

A different situation arises with the \tCV type constraints.
The \pCA pattern is the most frequently used to implement this type of constraint.
In our data, this pattern appears when the concrete value appears directly in the call to a setter or constructor.
It is important to note that instances of this pattern suffer from the ``magic number'' code smell \citep{Fowler2018}, which suggests that the use of this pattern is prone to introducing code smells.
An alternative is the less common \pAC pattern, which does not introduce this code smell.

Additionally, we find some uncommon patterns that apply to some particular situations and could possibly be adapted into the more common ones.
One example is the \pBFC pattern.
In this pattern, an integer is used as a bit field, which effectively turns it into an array of boolean values (a set of binary flags).
To check if one of the values is enabled, a \emph{mask} consisting of an integer constant with only the corresponding bit turned to 1 and the bitwise \emph{and} operation is applied with the value of the variable.
This kind of pattern is commonly found in languages such as C \citep{Oualline1997}.
The pattern could be converted into the \pBC pattern by turning each flag into its own boolean field.
The same idea is applicable to the \pNZC and \pNEC patterns.
These patterns exist because the \src{String} class has two possible empty values: empty string and \src{null}.
Hence, they could be turned into \pNC by ensuring that all empty strings are instead turned to \src{null} at creation.

Finally, we discuss the cases in which a constraint is implemented with a pattern that is the common implementation of a different constraint type.
For example, the constraint ``\textit{If m is less than zero}'' from Rhino \citep{rhino:ecma}, of \tVC type, is implemented with the \pBP pattern, when one would instead expect a \pBC pattern.
The enforcing statement is \src{if (sign[0])}, where \src{sign} is a boolean array of size 1.
This construction exists because this array is passed to another method that sets it according to the value of \src{m}, which is reminiscent of passing a parameter by reference in the C language \citep{Oualline1997}.
We argue that this is quite an unusual construction in Java, as passing by reference is not supported for primitive types such as boolean.
This unusual construction could be transformed into the more expected \pBC by having the method return the data instead of modifying it in-place.

Some constraints of \tDVC type are implemented with the \pBC pattern, and we attribute the implementation rationale to discrepancies between the language of the constraint and concrete implementation decisions.
For instance, the Ant constraint ``\textit{unless either the -verbose or -debug option is used}'' \citep{ant:targets} contains the constraint ``\textit{the -verbose option is used}'', which can be either true or false.
However, this is implemented by iterating over the arguments and checking each against the text of each option successively, shown in \listref{lst:at-verbose} (constraint implementation is line 6).
Such a long chain of \src{if--else} statements (more than 20 in this case) is a code smell \citep{Fard2013}.
Note that the contents of the \src{args} array could be cached into an object, which could later be queried on whether it contains the \emph{verbose} option, both getting rid of the code smell and applying the more common \pBP pattern.

\begin{lstlisting}[caption={Checking for verbose argument in Ant.},label=lst:at-verbose,float]
for (int i = 0; i < args.length; i++) {
  final String arg = args[i];
  [...]
  } else if (arg.equals("-quiet") || arg.equals("-q")) {
    msgOutputLevel = Project.MSG_WARN;
  } else if (arg.equals("-verbose") || arg.equals("-v")) {
    msgOutputLevel = Project.MSG_VERBOSE;
  } else if (arg.equals("-debug") || arg.equals("-d")) {
  [...]

\end{lstlisting}

\rqa{\textbf{RQ2 answer}} 
{We identified and defined \cips constraint implementation patterns. 
The \pBC and \pBP occur very frequently, while 13 other patterns are utilized frequently and the rest are rare, in our data.
This indicates that developers tend to use a rather small number patterns when implementing data constraints.
Additionally, our data suggest that certain patterns are commonly used to implement certain constraint types.
Finally, there is evidence that implementations of constraints that do not follow the most common patterns are due to unusual implementation decisions or exhibit code smells.
}

\section{Multiple Enforcements of a Constraint (RQ3)}
\label{sec:rq3}

As shown in the motivating example from \secref{sec:motivation}, constraint implementations may have multiple enforcing statements in different code locations.
We refer to them as being enforced in multiple distinct locations in the code or as having multiple enforcements.
Hence, such constraints have several trace links to the code (\ie one set of data definition statements and multiple distinct sets of enforcing statements).
Intuitively, one constraint should be enforced in one place in the code.
We study how many constraints are enforced in multiple locations for answering RQ3: \textit{What are the differences between multiple enforcing statements of the same constraint?}

The study for answering RQ2 relied on manually identifying only \textit{one} enforcing statement per constraint (and the corresponding data definitions).
For answering RQ3, we need to identify multiple enforcing statements for a given constraint, where they exist.
Unfortunately, it is prohibitively expensive to manually identify all enforcing statements of a constraint in large projects.
Hence, tool support for collecting additional traces is essential.
We leverage the constraint implementation patterns discovered in Section \ref{sec:rq2} and use static analysis techniques to automatically find candidate enforcing statements, based on the data definitions that were manually identified.

\subsection{Detectors for Tool-assisted Tracing}
\label{subsec:det_impl}

We implemented \dets static analysis-based detectors to assist the identification of multiple enforcing statements for a given constraint.
Each detector is designed to detect the instances of one frequent CIP (used in at least two instances in our data).
There are 15 frequent patterns in \tabref{tab:cip_distr}. 
We did not build detectors for two frequent patterns, \pPF (3 instances) and \pPM (2 instances).
The instances of the \pPF pattern do not appear in Java code, but in text files.
Those of the \pPM pattern make use of dynamic dispatch, hence static analysis may be insufficient for accurate detection.

Each detector uses the data definitions of a constraint as input and returns candidate enforcing statements.
The number of inputs that a detector accepts is the same as the number of  ``\textit{parts}'' of the pattern it implements, as defined in \tabref{tab:CIP}.
Hence, a detector may have 1 to 3 inputs depending on the pattern it implements.
For example, for detecting the \pBP pattern the detector takes a single operand as input.
For detecting the \pBFC the detector takes two operands as input, while for detecting the \pBC it uses two operands and one operator as input.

We use syntax analysis (for identifying the pattern) and dataflow analysis (for finding all instances of a pattern corresponding to a particular data definition) to automatically detect instances of our CIPs.
Syntax analysis at the Abstract Syntax Tree (AST) level is suitable for analyzing the source code structures, while the dataflow analysis is able to trace data dependence in an intermediate representation (IR).
Specifically, we implemented the detectors using a combination of \citet{JavaParser}---a parser with AST analysis capabilities, and \citet{wala}---a static analysis framework for Java.
For the AST analysis, we parse every Java file in the system's source code and record the lines where every instance of each pattern appears.
The instances are identified by matching code structures with \textit{statement type} and \textit{part} defined in \tabref{tab:CIP}.
For the WALA analysis, we first build a \textit{call graph} and a \textit{system dependence graph}, which is the program representation commonly used for program slicing \citep{Tip1994}.
For each of the CIPs, there exists data dependence between the data definitions and the enforcing statements.
We perform forward program slicing on the system dependence graph to track such data dependence.
In general, each detector performs slicing from the input data definitions and then matches any occurrences of the \textit{statement type} and \textit{part} defined in \tabref{tab:CIP} on the IR along the slice (or intersection of the slices, in case there are two operands).
It later confirms the match by checking that the source-code pattern exists in that location using the syntax analysis, as IR does not perfectly keep the code structures.

As an example, the constraint ``\textit{If [the buffer] has unsaved changes}'' \citep{jedit:closing} is enforced by the statement \src{if(buffer.isDirty())} with the \pBP pattern.
The data definition statement in this case is the definition of the \src{Buffer.dirty} field.
Passing this input to the \pBP detector returns a list of lines in files EditPane.java, View.java, BufferAutosaveRequest.java, among others, where the value of the field is used and the pattern appears.

\subsection{Tool-assisted Tracing Protocol}
Our goal is to retrieve trace links in addition to the ones identified manually in \secref{subsec:tracing}.
If a constraint is implemented using multiple enforcing statements, we create a separate trace link to each enforcing statement (and the associated data definitions). 
Because it is possible that multiple enforcing statements for the same constraint may follow different patterns, we use several detectors for each constraint.

Given a constraint, we execute all detectors that take the same number of inputs as the manually-traced pattern from \secref{subsec:patterns}.
Recall that the number of data definitions depends on the size of ``\textit{parts}" in the CIPs.
For example, if the manually-traced pattern of a constraint is \pNC pattern (which has a single part), we used its data definition to run all one-input detectors.
Therefore, these detectors would potentially find candidate enforcing statements that follow all patterns with a single part (\ie \pBP and \pNEC).

We used our detectors to retrieve candidate links for \tdcs constraints from all eight systems, \ie those implemented using one of the \dets patterns.
Two authors independently examined the candidate links for each constraint, classifying each link as true positive or false positive.
The authors followed the same protocol that was used to verify the traces in \secref{subsec:trace-validation}.
When the detectors returned more than 25 candidate links for a constraint, 25 of these were randomly sampled for classification.
In total, the authors inspected \mars candidate enforcing statements.

\subsection{Results and Analysis}
\label{subsec:validation}

We present a summary of the tool-assisted tracing results and analyze the cases where one constraint is enforced in multiple places.

\subsubsection{Tool-assisted Trace Validation}

We first verified that our detectors retrieved the manually-traced link for each constraint.
Retrieving a large number manually-traced links  would indicate that the detectors are suitable for this task.
At the same time, we expect and accept that some manually-traced links are not detected, as the consequence of the trade-off between performance, soundness, and precision in static analysis \citep{Livshits2015}.
The manually-traced links were retrieved for \tpgts out of \tdcs constraints (94\%). 
The detectors retrieved candidate links for \edscs out of \tdcs constraints, meaning that for 4 constraints, the detectors did not retrieve any candidate links.

On average, the detectors retrieved \arpc (median 4) candidate links per constraint.
After classifying the \mars candidate enforcing statements, our tool-assisted tracing identified \tps new links (\ie new enforcing statements).
Note that it is possible that the detectors retrieved some additional links which we did not classify, for 33 constraints for which the detectors retrieved more than 25 candidate links.
Nonetheless, the effort spent on retrieving and classify these new links is substantially less than the effort spent on the manual tracing.

\subsubsection{Analysis of the Multiple Enforcing Statements}

We analyzed the \edscs constraints for which the detectors returned at least one new enforcing statement.
For these constraints, we recovered a total of \edsis trace links, including \edscs manually and \tps using the detectors.
Among them, \cmi constraints (\pcmi of \edscs) have links to more than one enforcing statement, while the rest are implemented using a single enforcing statement.

\begin{figure}[htbp]
\begin{adjustbox}{addcode={\begin{minipage}{\width}}{\caption{Distribution of patterns for constraints with multiple enforcing statements. The five bars on the left correspond to the inconsistent constraints.}\label{fig:multi-impl} \end{minipage}},rotate=90,center}
	\includegraphics[width=\textheight]{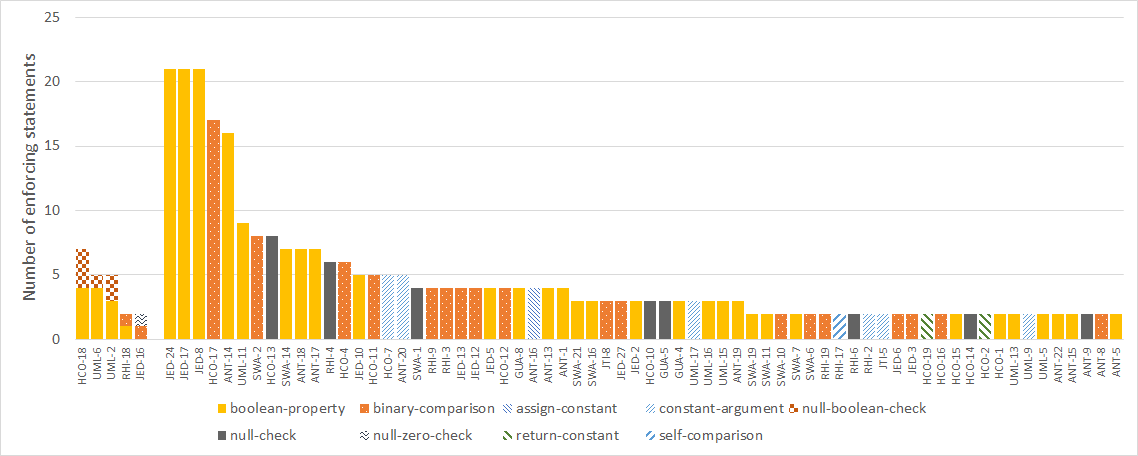}
\end{adjustbox}
\end{figure}

We further studied the enforcing statements of these \cmi constraints.
\figref{fig:multi-impl} 3 shows the distribution of the CIPs implementing these constraints.
We observed that in most cases (66 out of 71) the same pattern is used for all the enforcing statements of the same constraint (\ie the corresponding bar in \figref{fig:multi-impl} has a single texture and color).
We call them \emph{consistent} implementations.
In five cases, the constraint has more than one enforcing statement and they follow multiple patterns (\ie the corresponding bars in \figref{fig:multi-impl} show multiple textures and colors).
We call them \emph{inconsistent} implementations.

While these multiple enforcing statements are not inherently problematic, they are essentially instances of code cloning \citep{Baker1995}.
The cases of consistent implementations result in \emph{type 1} or \emph{type 2} clones, \ie they are either exact copies, or the only changes occur in identifiers and literals \citep{Bellon2007}.
However, inconsistent implementations lead to \emph{type 4} clones, in which the code is syntactically different, but the semantics are the same \citep{Roy2009}.

We identified two types of inconsistent implementations:

\begin{enumerate}
\item \textit{Related patterns.} The \textit{``supplied entity is already repeatable''} (HCO-18 in the chart) constraint for the HTTPComponents system \citep{http:spec} is checked using the method invocation \src{RequestEntityProxy .isRepeatable(request)} in four different code locations.
This repeated enforcing statement is an instance of the \pBP pattern. 
The constraint is additionally enforced in three different code locations using the enforcing statement: \src{entity != null \&\& entity.isRepeatable()}. 
This enforcing statement corresponds to the \pNBC pattern. 
These two patterns are similar, as they both check a boolean value, but the second one additionally accounts for a \src{null} value.
This is the case for the constraints HCO-18, UML-2, UML-6, and JED-16.

\item \textit{Unrelated patterns.} The Rhino constraint \textit{``[Result of toNumber] is [...] $+\infty$''} \citep{rhino:ecma} is implemented by an instance of \pBC pattern (\src{d == Double.POSITIVE\_INFINITY}) and an instance of \pBP pattern using a standard library utility method (\src{Double.isInfinite(d)}).
While, these two patterns are used to implement the same constraint, they have different structures.
This situation occurs in the implementation of constraint RHI-18.
\end{enumerate}

It is easy to argue that inconsistent implementations are detrimental to code maintainability, as \textit{type 4} clones are challenging to detect automatically \citep{Komondoor2001, Gabel2008}.
Additionally, they pose challenges when their rationale is not well documented (see example of related patterns above, the code does not specify why some cases require the \src{null} check while others do not).

Consistent implementations also pose potential problems.
Although existing research suggests that developers often evolve duplicated code consistently \citep{Thummalapenta2010}, handling a large number of duplicates (over 20 in some cases in our data) can lead to a more demanding and error-prone change process when these constraints need to be modified.
We argue that most of these enforcing statements can be refactored.
For example, the constraint SWA-1 ``\textit{configuration file is not
available}'' \citep{swarm:rg2} is implemented as \src{if (WinDataFile.configFile == null)}.
Refactoring this enforcing statement so that the null checking happens in a method of the \src{WinDataFile} would encapsulate the logic and make the semantics of the constraint clearer, which would make eventual changes easier.

The presence of duplicated code corresponding to business rules can also indicate the presence of duplicated business processes, which are challenging to identify in textual artifacts \citep{Guo2008}.

Even though the literature is divided on whether code clones are detrimental, evidenced by the extensive research on clone detection \citep{Ain2019, Roy2009}, or a necessary part of development \citep{Kapser2006}, we argue that it is counter-intuitive for data constraint implementations to exhibit a large amount of clones.

\rqa{\textbf{RQ3 answer}}
{Nearly half (44\%) of the constraints we studied are implemented with more than one enforcing statement.
These multiple enforcing statements result in code clones: 108 type 1, 138 type 2, and 10 type 4.
We attribute these implementations to design decisions, rather than to the intrinsic properties of the constraints.
We argue that most constraints with multiple enforcing statements would benefit from refactoring into simpler patterns, \ie \pBP.
}

\section{Threats to Validity and Limitations}
\label{sec:threats}

A major contribution of this paper is the discovery and definition of the constraint implementation patterns. 
These definitions are data-driven, created and agreed upon by the authors of this paper. 
It is possible that different coders would produce a different set of definitions. 
To mitigate this threat, we defined a clear coding framework based on qualitative data analysis methods \citep{Miles2014}, presented in \secref{subsec:patterns}. 
This process revealed 16 (9\%) cases of disagreement, which can be considered as a small proportion.

Our answer to RQ2 also depends on the accuracy of traces we produce in \secref{subsec:tracing}.
It is possible that some constraints could have been traced to the wrong statements in the code. 
To make our tracing as reliable as possible, we employed two tracers for each constraint and had two authors decide the final trace through a discussion.
This protocol is in line with previous work on traceability \citep{Eaddy2008a,Ali2011,Ali2013,Ali2012a}. 

Whether our CIP catalog reflects the space of constraint implementations in all Java systems depends on our choice of target systems. 
Since the systems in our data set are real-world open-source systems from a variety of domains, we expect the constraint implementations we identified to also exist in other similar systems.
Further research is needed to establish whether the CIPs and their distributions would be different in other type of software (\eg different domain and proprietary)
We make our data and our pattern catalog openly available, such that future research can enrich the catalog with new distribution information or with new patterns.

Our tool-assisted tracing protocol in \secref{sec:rq3} relied on the inputs and patterns that were derived from the manually defined traces, and it may not find all enforcing statements of each constraint.
The detectors can detect enforcing statements that are implemented using frequent CIPs, accounting for \freqpc of the constraint implementations in our data set.
Implementing additional detectors and using additional data is subject of future work.

Our answers to the RQs are also dependent on our choice to focus only on the data definition and enforcing statements. 
As we discussed in \secref{sec:motivation}, one can argue that there are other statements relevant to the implementation of a constraint, which should also be traced and analyzed.
For example, definitions of variables used by the enforcing statements.
We consider that the data definition and enforcing statements pair we trace to is a \textit{minimal subset} of a constraint implementation, that is, eliminating any of them would no longer produce non-ambiguous traces. 
We argue that including additional constructs in the traces will not alter or invalidate the current catalog of patterns.
Instead, it will likely result in the refinement of the existing patterns, based on the properties of these additional statements.
Expanding the study of constraint implementations to include additional code constructs/statements is subject of future work.

\section{Related Work}
\label{sec:related_work}

The concept of \textit{constraint} has been used in multiple contexts and with different definitions in the software engineering literature. 
We found two instances to be particularly related to our work.
In the context of business rules, Wiegers and Beatty \citep{Wiegers2013} define a constraint as ``a statement that restricts the actions that the system or its users are allowed to perform''.
This definition covers data constraints, because it can be said that the system \textit{is only allowed} to accept/produce valid data.
Breaux and Ant\'on \citep{Breaux2008} define a constraint as a statement that narrows the possible interpretations of a concept based on its properties.
For instance, ``\textit{patient who receives healthcare services}'' restricts the set of all patients to only those who receive healthcare services.
This definition is similar to ours in that data constraints conceptually narrow the set of possible entities, \eg \textit{``request to HEAD method''} is a subset of all requests.
It is important to note that neither of these works explore how the constraints are enforced in source code.

Previous studies have leveraged constraints in textual artifacts to extract usage patterns.
\citet{Xiao2012} use sentence patterns such as ``\textit{[noun] is allowed to [action] [resource]}'' to automatically extract security policies.
The extracted sentences correspond to constraints, though they constraint access control permissions, not data.
The work of \citet{Pandita2012} can infer data constraints (\eg ``\textit{path must not be null}''') from the documentation of a method (\eg the sentence ``\textit{If path is null}''). 
Similarly, other works classify method parameters according to whether a null value is allowed \citep{Tan2012}, has to belong to a specific type \citep{Zhou2017a}, or its numeric value has to be in a certain range \citep{Saied2015}.
Note that even though these techniques can infer the existence of a data constraint enforcement in the method, they do not study their implementations and depend on the accuracy of the documentation.

More similar to the study presented in this paper is the work of \citet{Yang2020}.
It examined three types of data constraints specific to web applications implemented in Ruby on Rails: front-end constraints 
expressed as regular expressions; application constraints on data fields in model classes, specified in validation functions (which check the length of a text-field, content uniqueness, and content presence); and database constraints specified in the applications'
migration files, through Rails Migration APIs.
The research found that these type of constraints are often checked inconsistently between the three architectural layers and developed a tool to identify such inconsistencies.
In contrast, we identified the constraints in textual documents and then manually traced them to their enforcing statements in Java system, regardless of their architecture or how they are implemented.
Our goal is identifying and analyzing all data constraint types and their implementations found in this systems, as opposed to restricting the set to a known type of implementation.

Research on automated business rule extraction proposed methods similar to those we used in the design of our detectors.
In particular, backward or forward slicing is done from a previously identified variable to detect the conditional statements that affect its value, hence related to a business rule \citep{Hatano2016, Cosentino2013, Huang1996, Sneed1996, Wang2004,Cosentino2012,Sneed2001, Chaparro2012}.
However, their goals were not to analyze implementation patterns of data constraints.

As part of our study, we performed manual and tool-assisted requirements-to-code traceability link recovery (RCTLR) \citep{Antoniol2002,Borg2014}, as we consider the data constraints as part of the system requirements.
We developed our own detectors for recovering candidate links, because existing approaches are not appropriate for this use.
This is because the current state of the art does not achieve retrieval of such links at statement level \citep{Cleland-Huang2014}.
Most existing techniques are based on text retrieval \citep{DeLucia2012,Borg2014}, while approaches based on machine learning \citep{Guo2017,Mirakhorli2016} and AI \citep{Sultanov2011,Blasco2020} have also been explored. 
Closer to our work are the approaches leveraging structural features of the software \citep{Eaddy2008,McMillan2009,Kuang2017}.
However, these features often focus on class or method relationships, and do not describe implementation patterns.
Recently, Blasco \etal \citep{Blasco2020} proposed a statement-level RCTLR approach that uses LSI and genetic algorithms.
It works by selecting a set of seed statements based on textual similarity, which is randomly mutated using the crossover and mutation operators until it results in a set of candidate links.
In contrast, our study found the statement-level candidate links by exploiting the implementation patterns that we identified.

The design and use of our detectors is related to research on the automated detection of design patterns. 
Three characteristics have mainly been used to identify patterns: structural 
\citep{Gueheneuc2008,Tsantalis2006,Kaczor2006,Gueheneuc2004},
behavioral \citep{Shi2006,Park2004}, 
and semantic \citep{Dong2007}.
Our work is similar to those that employ structural characteristics, as we use static analysis to pinpoint the location where a certain constraint is enforced. 
However, our patterns span only statements, while design patterns span multiple classes, focusing on more generic computational solutions for recurring problems.

\section{Conclusions and Future Work}
\label{sec:conclusions}

While the importance of business rules is widely recognized in software engineering and the field of automated business rule extraction provides a wealth of techniques, there still is a lack of understanding of how business rules are implemented in source code.
This is not surprising, given the vast diversity in possible rules and implementation decisions.
This study is a first step towards better understanding how developers implement business rules.

We focused on understanding data constraints and their implementations through an empirical study.
Studying \tcs constraints from eight Java systems, we learned that:

\begin{enumerate}
\item The studied systems describe four types of data constraints.

\item The implementations of the \tcsne studied data constraints (those not enforced externally) can be categorized into \cips constraint implementation patterns (CIPs).
\freqcips of these patterns (\freqcipspc) implement \freqpcimp (\freqpc) of the constraints, with the two most common (\pBP and \pBC) accounting for half of all the implementations.
This suggests that developers employ a small number of CIPs to implement most constraints.

\item Certain patterns are preferred when implementing constraints of certain types, and that deviations from these trends are associated with unusual implementation decisions and code smells.

\item 44\% of the studied constraints are implemented with more than one enforcing statement in multiple code locations. 
While 93\% of them use the same pattern for all of the enforcing statements, they exhibit code cloning (\ie type 1, type 2, and type 4 clones).
\end{enumerate}

We expect that these insights will enable the following research avenues:

\begin{enumerate}
\item The catalog of CIPs is an evolving contribution, which can always be extended by examining new systems.
Specifically, identification of CIPs in other programming languages is part of our future work.

\item The relationship between constraint types and CIPs can enable the creation of best practice guidelines.

\item The CIPs and our automated detectors can be used to improve existing automated business rules extraction approaches and traceability link recovery tools.

\end{enumerate}

\bibliographystyle{spbasic}      
\bibliography{artifacts,references}

\appendix
\appendixpage
\section{Constraint Implementation Patterns Catalog}

Tables \ref{tab:CIP2}, \ref{tab:CIP3}, and \ref{tab:CIP4} contain 23 descriptions of the constraint implementation patterns from our catalog.
The 7 most common patterns can be found on \tabref{tab:CIP}.

\renewcommand{\arraystretch}{0.92}
\begin{table*}[htbp]
			
	\caption{CIP catalog, part 2.}
	\label{tab:CIP2}%
	
	\small
	
	\begin{tabular}{p{\textwidth}}
		
		\toprule
		
		\textbf{CIP name:} \pEOC. \\
		\textbf{Description:} Equality expressions (or \src{equals} method calls) are chained by ``or'' operators in an expression checking possible values of a variable.\\
		\textbf{Statement type:} Boolean expression. \\
		\textbf{Parts:} $\{\mathit{variable}\}$ \\ 
		\textbf{Example:} \\ 
		\quad \textit{Instance:} \srcf{option.equals(\textquotedblleft true\textquotedblright) || option.equals(\textquotedblleft on\textquotedblright) || option.equals(\textquotedblleft yes\textquotedblright)} \\
		\quad \textit{Parts:} $\{$\srcf{option}$\}$.\\
		
		\midrule
		
		\textbf{CIP name:} \pPF. \\
		\textbf{Description:} The value for a variable is stored in a file. \\
		\textbf{Statement type:} File line. \\
		\textbf{Parts:} $\{\mathit{constant}\}$ \\ 
		\textbf{Example:} \\ 
		\quad \textit{Instance:} \srcf{backups=1} \\
		\quad \textit{Parts:} $\{$\srcf{backups}$\}$.\\
		\midrule	

			\textbf{CIP name:} \pPM. \\
		\textbf{Description:} Conditional branching is achieved by calling a method in a superclass that is overridden in a subclass.\\
		\textbf{Statement type:} Method call. \\
		\textbf{Parts:} $\{\mathit{method}\}$ \\ 
		\textbf{Example:} \\ 
		\quad \textit{Instance:} \srcf{scriptable.getDefaultValue()}  (\srcf{getDefaultValue} is an abstract method)\\
		\quad \textit{Parts:} $\{$\srcf{Scriptable.getDefaultValue()}$\}$.\\
		\midrule	
		
		\textbf{CIP name:} \pNEC. \\
		\textbf{Description:} A string value is checked for nullity using the \texttt{\footnotesize ==} or \texttt{\footnotesize !=} operators and then compared to empty string using the equals method. The first expression is a \pNC pattern, but for \pNEC to apply, both expressions must be present.\\
		\textbf{Statement type:} Boolean expression. \\
		\textbf{Parts:} $\{\mathit{variable}\}$ \\ 
		\textbf{Example:} \\ 
		\quad \textit{Instance:} \srcf{string == null || string.equals(\textquotedblleft \textquotedblright) }  \\
		\quad \textit{Parts:} $\{$\srcf{string}$\}$.\\
		
		\midrule	
		
		\textbf{CIP name:} \pNZC. \\
		\textbf{Description:} A value is checked for nullity using the \texttt{\footnotesize ==} or  \texttt{\footnotesize !=} operators and then its length or other numeric property is compared to zero. The first expression is a \pNC pattern, but for \pNZC to apply, both expressions must be present.\\
		\textbf{Statement type:} Boolean expression. \\
		\textbf{Parts:} $\{\mathit{variable}\}$ \\ 
		\textbf{Example:} \\ 
		\quad \textit{Instance:} \srcf{string != null \&\& string.length() > 0} \\
		\quad \textit{Parts:} $\{$\srcf{string}$\}$.\\
		
		\midrule	
		
		\textbf{CIP name:} \pRC. \\
		\textbf{Description:} Return a literal value. \\
		\textbf{Statement type:} Return statement. \\
		\textbf{Parts:} $\{\mathit{constant}\}$ \\ 
		\textbf{Example:} \\ 
		\quad \textit{Instance:} \srcf{return 80} \\
		\quad \textit{Parts:} $\{$\srcf{80}$\}$.\\
		
		\midrule	
		
		\textbf{CIP name:} \pSLC. \\
		\textbf{Description:} A switch is done first on the length of a string and then on specific characters to determine which of the options corresponds to the input string.\\
		\textbf{Statement type:} Switch statement. \\
		\textbf{Parts:} $\{\mathit{variable}\}$ \\ 
		\textbf{Example:} \\ 
		\quad \textit{Instance:} \srcf{switch(token.length()) \{case 1: c=s.charAt(1); if (c=='f') \{ $\ldots$ \} $\ldots$}  \\
		\quad \textit{Parts:} $\{$\srcf{token}$\}$.\\
		\midrule	
		
		\textbf{CIP name:} \pSC. \\
		\textbf{Description:} A variable is compared to itself.\\
		\textbf{Statement type:} Relational expression. \\
		\textbf{Parts:} $\{\mathit{variable}\}$ \\ 
		\textbf{Example:} \\ 
		\quad \textit{Instance:} \srcf{d != d} \\
		\quad \textit{Parts:} $\{$\srcf{d}$\}$.\\

		\bottomrule
	\end{tabular}
\end{table*}

\renewcommand{\arraystretch}{0.92}
\begin{table*}[htbp]
			
	\caption{CIP catalog, part 3.}
	\label{tab:CIP3}%
	
	\small
	
	\begin{tabular}{p{\textwidth}}
		
		\toprule
		
		\textbf{CIP name:} \pSS. \\
		\textbf{Description:} The \src{startsWith} method is called on a string variable.\\
		\textbf{Statement type:} Method call. \\
		\textbf{Parts:} $\{\mathit{variable}\}$ \\ 
		\textbf{Example:} \\ 
		\quad \textit{Instance:} \srcf{arg.startsWith("-background")} \\
		\quad \textit{Parts:} $\{$\srcf{arg}$\}$.\\
		
		\midrule
		
		\textbf{CIP name:} \pNBC. \\
		\textbf{Description:} A variable is checked for nullity using the \texttt{\footnotesize ==} or \texttt{\footnotesize !=} operators and then a boolean property of the variable is checked. The first expression is a \pNC pattern, and the second is a \pBP, but for \pNBC to apply, both expressions must be present. \\
		\textbf{Statement type:} Boolean expression. \\
		\textbf{Parts:} $\{\mathit{variable}\}$ \\ 
		\textbf{Example:} \\ 
		\quad \textit{Instance:} \srcf{saveAction != null \&\& saveAction.isEnabled()} \\
		\quad \textit{Parts:} $\{$\srcf{saveAction}$\}$.\\
		
		\midrule
		
		\textbf{CIP name:} \pS. \\
		\textbf{Description:} A setter method is used to assign a value to a field. \\
		\textbf{Statement type:} Method call. \\
		\textbf{Parts:} $\{\mathit{method}, \mathit{variable}\}$ \\ 
		\textbf{Example:} \\ 
		\quad \textit{Instance:} \srcf{project.setBasedir(helperImpl.buildFileParent.getAbsolutePath())} \\
		\quad \textit{Parts:} $\{$\srcf{project.setBasedir}, \srcf{helperImpl.buildFileParent.getAbsolutePath()}$\}$.\\
		
		\midrule
		
		\textbf{CIP name:} \pCAs. \\
		\textbf{Description:} A field is initialized in a constructor or builder method, but not using any of the parameters. \\
		\textbf{Statement type:} Assignment. \\
		\textbf{Parts:} $\{\mathit{field}\}$ \\ 
		\textbf{Example:} \\ 
		\quad \textit{Instance:} \srcf{authorname = Configuration.getString(Argo.KEY\_USER\_FULLNAME)} \\
		\quad \textit{Parts:} $\{$\srcf{authorname}$\}$.\\
		
		\midrule
		
		\textbf{CIP name:} \pDC. \\
		\textbf{Description:} Two variables are subtracted and their difference is compared to zero. \\
		\textbf{Statement type:} Arithmetic expression, Boolean expression. \\
		\textbf{Parts:} $\{\mathit{variable}, \mathit{variable}\}$ \\ 
		\textbf{Example:} \\ 
		\quad \textit{Instance:} \srcf{int delta = getMajor() - that.getMajor(); if (delta == 0);} \\
		\quad \textit{Parts:} $\{$\srcf{getMajor}, \srcf{getMajor}$\}$.\\
		
		\midrule
		
		\textbf{CIP name:} \pEV. \\
		\textbf{Description:} The method valueOf of an enum is used to ensure that a string variable represents a valid member of the enum. \\
		\textbf{Statement type:} Method call. \\
		\textbf{Parts:} $\{\mathit{variable}\}$ \\ 
		\textbf{Example:} \\ 
		\quad \textit{Instance:} \srcf{BufferSet.Scope.valueOf(jEdit.getProperty("bufferset.scope", "global"))} \\
		\quad \textit{Parts:} $\{$\srcf{jEdit.getProperty("bufferset.scope", "global")}$\}$.\\
		
		\midrule
		
		\textbf{CIP name:} \pICL. \\
		\textbf{Description:} The value of the variable is checked by iterating over a collection of possible values and checking equality for each one. The value of this collection comes from a literal. \\
		\textbf{Statement type:} Loop statement. \\
		\textbf{Parts:} $\{\mathit{variable}, \mathit{collection}\}$ \\ 
		\textbf{Example:} \\ 
		\quad \textit{Instance:} \srcf{for (ExtensionType value : values) \{if (name.equals(value.name())) ... } \\
		\quad \textit{Parts:} $\{$\srcf{name}, \srcf{values}$\}$.\\
		
		\midrule
		
		\textbf{CIP name:} \pMO. \\
		\textbf{Description:} Restricts the values that a variable can take to the possible remainders of a division. \\
		\textbf{Statement type:} Arithmetic expression. \\
		\textbf{Parts:} $\{\mathit{variable}\}$ \\ 
		\textbf{Example:} \\ 
		\quad \textit{Instance:} \srcf{daysSince19700101 \% 7} \\
		\quad \textit{Parts:} $\{$\srcf{daysSince19700101}$\}$.\\
		
		\bottomrule
	\end{tabular}
\end{table*}

\renewcommand{\arraystretch}{0.92}
\begin{table*}[htbp]
			
	\caption{CIP catalog, part 4.}
	\label{tab:CIP4}%
	
	\small
	
	\begin{tabular}{p{\textwidth}}
		
		\toprule
		
		\textbf{CIP name:} \pSE. \\
		\textbf{Description:} The endsWith method is called on a string variable.\\
		\textbf{Statement type:} Method call. \\
		\textbf{Parts:} $\{\mathit{variable}\}$ \\ 
		\textbf{Example:} \\ 
		\quad \textit{Instance:} \srcf{name.toLowerCase().endsWith("." + defaultFilter.getSuffix())} \\
		\quad \textit{Parts:} $\{$\srcf{name}$\}$.\\
		
		\midrule
		
		\textbf{CIP name:} \pSwC. \\
		\textbf{Description:} One of the cases of the switch checks the value (switch variable is of type enum). \\
		\textbf{Statement type:} Switch statement. \\
		\textbf{Parts:} $\{\mathit{variable}\}$ \\ 
		\textbf{Example:} \\ 
		\quad \textit{Instance:} \srcf{switch(state) \{case Buffer.FILE\_CHANGED: \ldots} \\
		\quad \textit{Parts:} $\{$\srcf{state}$\}$.\\
		
		\midrule
		
		\textbf{CIP name:} \pOVS. \\
		\textbf{Description:} Each allowable value for a set is defined as the return value of the override of an abstract method. \\
		\textbf{Statement type:} Method definition. \\
		\textbf{Parts:} $\{\mathit{method}\}$ \\ 
		\textbf{Example:} \\ 
		\quad \textit{Instance:} \srcf{public abstract String getExtension();} \\
		\quad \textit{Parts:} $\{$\srcf{getExtension}$\}$.\\
		
		\midrule
		
		\textbf{CIP name:} \pCSC. \\
		\textbf{Description:} A numeric variable is cast to another type and then compared to the original variable. \\
		\textbf{Statement type:} Assignment, Boolean expression. \\
		\textbf{Parts:} $\{\mathit{variable}\}$ \\ 
		\textbf{Example:} \\ 
		\quad \textit{Instance:} \srcf{int id = (int)d; if (id == d)} \\
		\quad \textit{Parts:} $\{$\srcf{d}$\}$.\\
		
		\midrule
		
		\textbf{CIP name:} \pILF. \\
		\textbf{Description:} Iterate over collection of possible values. If the variable matches at some point, return the index. Otherwise return -1 at the end. \\
		\textbf{Statement type:} Loop statement, Return statement. \\
		\textbf{Parts:} $\{\mathit{collection}, \mathit{variable}\}$ \\ 
		\textbf{Example:} \\ 
		\quad \textit{Instance:} \srcf{for (int i = 0; i < values.length; i++) \{if (value.equals(values[i])) \{return i;\}\} return -1;} \\
		\quad \textit{Parts:} $\{$\srcf{values, value}$\}$.\\
		
		\midrule
		
		\textbf{CIP name:} \pACC. \\
		\textbf{Description:} Assigns a value derived from a method call on the result of a \src{.class} construct. \\
		\textbf{Statement type:} Assignment. \\
		\textbf{Parts:} $\{\mathit{variable}\}$ \\ 
		\textbf{Example:} \\ 
		\quad \textit{Instance:} \srcf{classname = DefaultExecutor.class.getName();} \\
		\quad \textit{Parts:} $\{$\srcf{classname}$\}$.\\
		
		\midrule
		
		\textbf{CIP name:} \pIRC. \\
		\textbf{Description:} A chain of ifs is used like a switch on a field, checking against the possible values of the variable. There are no else blocks and the body of each \src{if} is a return statement. \\
		\textbf{Statement type:} If statement. \\
		\textbf{Parts:} $\{\mathit{variable}\}$ \\ 
		\textbf{Example:} \\ 
		\quad \textit{Instance:} \srcf{if ("jikes".equalsIgnoreCase(compilerType)) \{return new Jikes();\} if ("extjavac".equalsIgnoreCase(compilerType)) \{return new JavacExternal();\} \ldots} \\
		\quad \textit{Parts:} $\{$\srcf{compilerType}$\}$.\\
		
		\bottomrule
	\end{tabular}
\end{table*}

\end{document}